\newcommand{\Tr}{\mathop{\mathrm{Tr}} \nolimits}

\documentclass[prb,twocolumn,amsmath,amssymb,superscriptaddress]{revtex4-1}
\usepackage{graphicx,bm,color,mathptmx,hyperref}

\newcommand{\ket}[1]{|{#1}\rangle}
\newcommand{\bra}[1]{\langle #1|}

\newcommand{\I}{\mathrm{i}}
\newcommand{\E}{\mathrm{e}}
\newcommand{\braket}[2]{\langle #1 | #2 \rangle}

\begin{document}

\title{Wavefront sensing reveals optical coherence}

\author{B. Stoklasa} 
\affiliation{Department of Optics, Palacky University, 
17. listopadu 12, 771 46 Olomouc, Czech Republic}

\author{L. Motka} 
\affiliation{Department of Optics, Palacky  University, 
17. listopadu 12, 771 46 Olomouc, Czech Republic}

\author{J. Rehacek} 
\affiliation{Department of Optics, Palacky University, 
17. listopadu 12, 771 46 Olomouc, Czech Republic}

\author{Z. Hradil} 
\affiliation{Department of Optics, Palacky University, 
17. listopadu 12, 771 46 Olomouc, Czech Republic}

\author{L. L. S\'{a}nchez-Soto} 
 \affiliation{Departamento de \'Optica, Facultad de F\'{\i}sica, 
 Universidad Complutense, 28040~Madrid,  Spain}
\email[Correspondence and requests for materials should be addressed
to  L.L.S.S. e-mail: ]{lsanchez@fis.ucm.es}

\begin{abstract}
  Wavefront sensing is a set of techniques providing efficient means
  to ascertain the shape of an optical wavefront or its deviation from
  an ideal reference. Due to its wide dynamical range and high optical
  efficiency, the Shack-Hartmann is nowadays the most widely used of
  these sensors.  Here, we show that it actually performs a
  simultaneous measurement of position and angular spectrum of the
  incident radiation and, therefore, when combined with tomographic
  techniques previously developed for quantum information processing,
  the Shack-Hartmann can be instrumental in reconstructing the
  complete coherence properties of the signal.  We confirm these
  predictions with an experimental characterization of partially
  coherent vortex beams, a case that cannot be treated with the
  standard tools.  This seems to indicate that classical methods
  employed hitherto do not fully exploit the potential of the
  registered data.
\end{abstract}

\maketitle

Light is a major carrier of information about the universe around us,
from the smallest to the largest scale. Three-dimensional objects emit
radiation that can be viewed as complex wavefronts shaped by diverse
features, such as refractive index, density, or temperature of the
emitter. These wavefronts are specified by both their amplitude and
phase; yet, as conventional optical detectors measure only
(time-averaged) intensity, information on the phase is discarded.
This information turns out to be valuable for a variety of
applications, such as optical testing~\cite{Malacara:2007kx}, image
recovery~\cite{Dai:2008zr}, displacement and position
sensing~\cite{Ares:2000fk}, beam control and
shaping~\cite{Katz:2011ly,McCabe:2011ij,Mosk:2012bs}, as well as
active and adaptive control of optical systems~\cite{Tyson:2011ys}, to
mention but a few.

Actually, there exists a diversity of methods for wavefront
reconstruction, each one with its own pros and
cons~\cite{Geary:1995fk}.  Such methods can be roughly classified into
three categories: (a) interferometric methods, based on the
superposition of two beams with a well-defined relative phase; (b)
methods based on the measurement of the wavefront slope or wavefront
curvature, and (c) methods based on the acquisition of images followed
by the application of an iterative phase-retrieval
algorithm~\cite{Luke:2002ad}.  Notwithstanding the enormous progress
that has already been made, practical and robust wavefront sensing
still stands as an unresolved and demanding
problem~\cite{Campbell:2006if}.

The time-honored example of the Shack-Hartmann (SH) wavefront sensor
surely deserves a special mention~\cite{Platt:2001dz}: its wide
dynamical range, high optical efficiency, white light capability, and
ability to use continuous or pulsed sources make of this setup an
excellent solution in numerous applications.

The operation of the SH sensor appeals to the intuition, giving the
overall impression that the underlying theory is
obvious~\cite{Primot:2003fe}. Indeed, it is often understood in an
oversimplified geometrical-optics framework, which is much the same as
assuming full coherence of the detected signal.  By any means, this is
not a complete picture: even in the simplest instance of beam
propagation, the coherence features turn out to be
indispensable~\cite{Mandel:1995qy}.

It has been recently suggested~\cite{Hradil:2010fv} that SH sensing
can be reformulated in a concise quantum notation.  This is more than
an academic curiosity, because it immediately calls for the
application of the methods of quantum state
reconstruction~\cite{lnp:2004uq}. Accordingly, one can verify right
away that wavefront sensors may open the door to an assessment of the
mutual coherence function, which conveys full information on the
signal.

In this paper, we report the first experimental measurement of the
coherence properties of an optical beam with a SH sensor. To that end,
we have prepared several coherent and incoherent superpositions of
vortex beams. Our strategy can efficiently disclose that information,
whereas the common SH operation fails in the task.

\begin{figure}[b]
  \centerline{\includegraphics[width=0.71 \columnwidth]{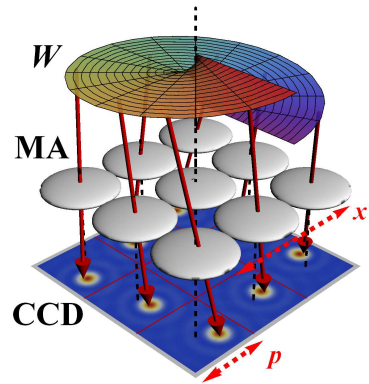}}
  \caption{\textbf{The principle of the SH wavefront sensor.} A microlens array
    (MA) subdivides the wavefront ($W$) into multiple beams that are
    focused in a CCD camera. Local slope of the wavefront over each
    microlens aperture determines the location of the spot on the
    CCD. Red arrows represent normals to the wavefront.}
  \label{figSH}
\end{figure}

\bigskip

\noindent{\textbf{Results}}

\noindent{\textbf{SH wavefront sensing.}}
The working principle of the SH wavefront sensor can be elaborated
with reference to Fig.~\ref{figSH}. An incoming light field is divided
into a number of sub-apertures by a microlens array that creates focal
spots, registered in a CCD camera. The deviation of the spot pattern
from a reference measurement allows the local direction angles to be
derived, which in turn enables the reconstruction of the wavefront. In
addition, the intensity distribution within the detector plane can be
obtained by integration and interpolation between the foci.

Unfortunately, this naive picture breaks down when the light is
partially coherent, because the very notion of a single wavefront
becomes somewhat ambiguous: the signal has to be conceived as a
statistical mixture of many wavefronts~\cite{Goodman:2005qa}. To
circumvent this difficulty, we observe that these sensors provide a
simultaneous detection of position and angular spectrum (i.e.,
directions) of the incident radiation.  In other words, the SH is a
pertinent example of a simultaneous unsharp position and momentum
measurement, a question of fundamental importance in quantum theory
and about which much has been
discussed~\cite{Arthurs:1965pi,Stenholm:1992lh,Raymer:1994ye}.

Rephrasing the SH operation in a quantum parlance will prove pivotal
for the remaining discussion.  Let $\varrho$ be the coherence matrix
of the field to be analyzed. Using an obvious Dirac notation, we can
write $ G ( \mathbf{x}^{\prime}, \mathbf{x}^{\prime \prime} ) =
\langle \mathbf{x}^{\prime}| \varrho | \mathbf{x}^{\prime \prime}
\rangle= \Tr ( \varrho |\mathbf{x}^{\prime} \rangle \langle
\mathbf{x}^{\prime \prime}|)$, where $|\mathbf{x}\rangle$ is a vector
describing a point-like source located at $\mathbf{x}$ and $\Tr$ is
the matrix trace. Thereby, the mutual coherence function $G (
\mathbf{x}^{\prime}, \mathbf{x}^{\prime \prime} )$ appears as the
position representation of the coherence matrix.  As a special case,
the intensity distribution across a transversal plane becomes $I(
\mathbf{x} )= \Tr ( \varrho | \mathbf{x} \rangle \langle
\mathbf{x}|)$.  Moreover, a coherent beam of complex amplitude $U(
\mathbf{x})$, can be assigned to a ket $\ket{U}$, such that $U(
\mathbf{x} )=\langle \mathbf{x} | U \rangle$.

To simplify, we restrict the discussion to one dimension, denoted by
$x$.  If the setup is illuminated with a coherent signal $U(x)$, and
the $i$th microlens is $\Delta x_{i}$ apart from the SH axis, this
microlens feels the field $U( x-\Delta x_{i} ) = \bra{x} \exp(-i
\Delta x_{i} \, P ) \ket{U}$, where $P$ is the momentum operator. This
field is truncated and filtered by the aperture (or pupil) function
$A(x)=\langle x| A\rangle$ and Fourier transformed by the microlens
prior to being detected by the CCD camera. All this can be accounted
for in the form
\begin{equation}
  U^{\prime} (\Delta p_{j}) = \bra{A} \exp (-i \Delta p_{i}  \, X) 
  \exp (-i \Delta x_i \, P) \ket{U} \, ,
\end{equation}
where $X$ is the position operator and we have assumed that the $j$th
pixel is angularly displaced from the axis by $\Delta p_{j}$.  The
intensity measured at the $j$th pixel behind the $i$th lens is then
governed by a Born-like rule
\begin{equation}
  \label{born}
  I (\Delta x_{i},\Delta p_{j} ) =  
  \Tr ( \varrho \ket{\pi_{ij}} \bra{\pi_{ij}} ) \, , 
\end{equation}
with $\ket{\pi_{ij}}= \exp ( i\Delta x_i \, P) \exp (i \Delta p_i \,
X) \ket{A}$.  As a result, each pixel performs a projection on the
position- and momentum-displaced aperture state, as anticipated
before.

Some special cases of those aperture states are particularly
appealing. For pointlike microlenses, $A(x) \rightarrow \delta(x)$ and
$\ket{\pi_{ij}} \rightarrow \ket{x=\Delta x_{i}}$ (i.e., a position
eigenstate): they produce broad diffraction patterns and information
about the transversal momentum is lost.  Conversely, for very large
microlenses, $A(x) \rightarrow 1$ and $\ket{\pi_{ij}} \rightarrow
\ket{p=\Delta p_j}$ (i.e., a momentum eigenstate): they provide a
sharp momentum measurement with the corresponding loss of position
sensitivity.  A most interesting situation is when one uses a Gaussian
approximation~{\cite{Hradil:2010fv}; now $A(x)=\exp(-x^2/2)$, which
implies $\ket{\pi_{ij}} \rightarrow \ket{\alpha_{ij}}$, that is, a 
coherent state of amplitude $\alpha_{ij}=\Delta x_{i} + i \Delta
p_{j}$. This means that the measurement in this case projects the
signal on a set of coherent states and hence yields a direct  sampling
of the Husimi distribution~\cite{Husimi:1940fu}
$Q(\alpha)=\bra{\alpha}\varrho\ket{\alpha}$. 

This quantum analogy provides quite a convenient description of the
signal: different choices of CCD pixels and/or microlenses can be
interpreted as particular phase-space
operations~\cite{Lvovsky:2009ys}.

\bigskip

\noindent{\textbf{SH tomography. }}
Unlike the Gaussian profiles discussed before, in a realistic setup
the microlens apertures do not overlap. If we introduce the operators
$\Pi_{ij}=\ket{\pi_{ij}}\bra{\pi_{ij}}$, the measurements describing
two pixels belonging to distinct apertures are compatible whenever
$[\Pi_{ij},\Pi_{i'j}]=0$, $i\neq i^{\prime}$, which renders the scheme
informationally incomplete~\cite{Busch:1989gb}. Signal components
passing through distinct apertures are never recombined and the mutual
coherence of those components cannot be determined.

Put differently, the method cannot discriminate signals comprised of
sharply-localized non-overlapping components. Nevertheless, these
problematic modes do not set any practical restriction. As a matter of
fact, spatially bounded modes (i.e., with vanishing amplitude outside
a finite area) have unbounded Fourier spectrum and so, an unlimited
range of transversal momenta. Such modes cannot thus be prepared with
finite resources and they must be excluded from our considerations:
for all practical purposes, the SH performs an informationally
complete measurement and any practically realizable signal can be
characterized with the present approach.

\begin{figure*}
  \centering{\includegraphics[scale=0.7]{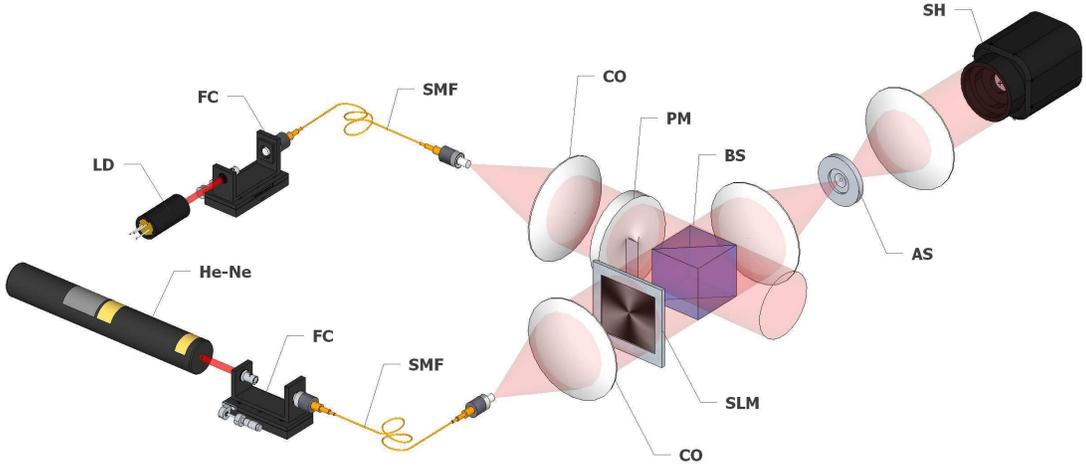}}
  \caption{\textbf{Experimental layout for preparing and detecting partially
    coherent vortex beams.}  Two independent laser sources, He-Ne at
    633~nm (He-Ne) and a laser diode at 635~nm (LD), are coupled into
    single-mode fibers (SMF) by fiber couplers (FC). After collimation
    (CO) , they are transformed into vortex beams by two different
    techniques. The first beam, representing a coherent superposition
    of two vortex modes, is prepared by a digital hologram imprinted
    in a spatial light modulator (SLM). Unwanted diffraction orders
    are filtered by an aperture stop (AS), placed in a 4$f$
    system. The second beam is modulated by a vortex phase mask (PM)
    and represents a single vortex mode with opposite phase respect to
    the first beam. Both beams are incoherently mixed in a beam
    splitter (BS) and finally detected in a SH sensor (SH).}
  \label{figSetup}
\end{figure*}

To proceed further in this matter, we expand the signal as a finite
superposition of a suitable spatially-unbounded computational basis
(depending on the actual experiment, one should use plane waves,
Laguerre-Gauss beams, etc).  If that basis is labeled by $\ket{k}$
($k=1,\ldots,d$, with $d$ being the dimension), the complex amplitudes
are $\braket{x}{k}=\psi_{k} (x)$.  Therefore, the coherence matrix
$\varrho$ and the measurement operators $\Pi_{ij}$ are given by
$d\times d$ non-negative matrices. A convenient representation of
$\Pi_{ij}$ can be obtained directly from Eq.~\eqref{born}, viz,
\begin{equation}
  \label{matrixel}
  \left(\Pi_{ij}\right)_{m n} =\psi_{n,i} (\Delta p_j) \, \psi_{m,i}^\ast(\Delta p_j),
\end{equation}
where $\psi_{m,i}(x)$ is the complex amplitude at the CCD plane of the
$i$th lens generated by the incident $m$th basis mode $\psi_{m}$.

This idea can be illustrated with the simple yet relevant example of
square microlenses: $A(x)=\mathrm{rect}(x)$. We decompose the signal
in a discrete set of plane waves $\psi_k(x)=\exp(-i p_k x)$,
parametrized by the transverse momenta $p_k$.  This is just the
Fraunhofer diffraction on a slit, and the measurement matrix is
\begin{equation}
  \label{square}
  \left(\Pi_{ij}\right)_{m n} = \mathrm{sinc}(\Delta p_{j}+p_{m}) 
  \mathrm{sinc}( \Delta p_{j} +p_{n} )
  e^{i (p_{m} -p_{n}) \Delta  x_{i}} \, .
\end{equation}
The smallest possible search space consists of two plane waves (which
is equivalent to a single-qubit tomography).  By considering different
pixels $j$ belonging to the same aperture $i$, linear combinations of
only three out of the four Pauli matrices can be generated from
Eq.~\eqref{square}.  For example, a lens placed on the SH axis
($\Delta x_i=0$) fails to generate $\sigma_y$ and at least one more
lens with a different $\Delta x_i$ needs to be added to the setup to
make the tomography complete.

This argument can be easily extended: the larger the search space, the
more microlenses must be used.  In this example, the maximum number of
independent measurements generated by the SH detection is
$(2M+1)d-3M$, for $M$ lenses.  A $d$-dimensional signal ---a spatial
qudit--- can be characterized with about $M \sim d/2$
microlenses. This should be compared to the $ d$ quadratures required
for the homodyne reconstruction of a photonic
qudit~\cite{Leonhardt:1996bh,Sych:2012qo}.

\bigskip

\noindent{\textbf{Experiment.}}
We have validated our method with vortex
beams~\cite{Molina:2007kn,Torres:2011vn}.  Consider the one-parameter
family of modes specified by the orbital angular momentum $\ell$,
$V_{\ell} = \langle r,\varphi|\text{V}_{\ell} \rangle\propto \E^{\I
  \ell \varphi} $, where $(r, \varphi)$ are cylindrical coordinates.
In our experiment, the partially coherent signal
\begin{equation}
  \label{eq:true}
  \varrho_{\text{true}}= 
  |V_{-3} - \textstyle{\frac{i}{2}} V_{-6} \rangle  
  \langle V_{-3} - \textstyle{\frac{i}{2}}   V_{-6} | + 
  \frac{1}{2} |V_3 \rangle \langle V_3| 
\end{equation}
was created; that is, modes $V_{-3}$ and $V_{-6}$ are coherently
superposed, while $V_{3}$ is incoherently mixed. Figure~\ref{figSetup}
sketches the experimental layout used to generate~(\ref{eq:true}).
Imperfections of the setup and sensor noise makes the actual state to
differ from the true state.  Calibration and signal intensity scans
are presented in Fig.~\ref{figData}.

The coherence matrix of the true state is expanded in the
7-dimensional space spanned by the modes $V_{\ell}$, with $\ell \in \{
-9, -6, -3, 0, +3, +6, +9\}$. The resulting matrix elements are
plotted in Fig.~\ref{figQ}.

\begin{figure}
  \centerline{\includegraphics[width=\columnwidth]{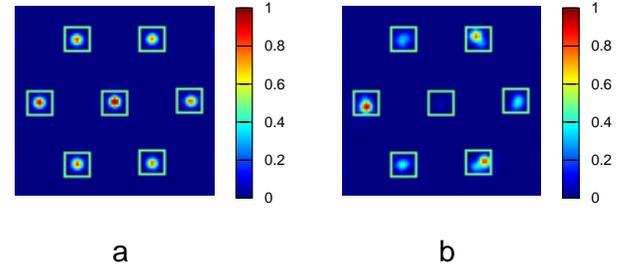}}
  \caption{\textbf{Experimental CCD signal.}  Rescaled 8-bit data
    corresponding to 7 microlenses placed in a hexagonal geometry, $81
    \times 81$-pixels region, is displayed in both panels. a) Data of
    the plane wave used for calibration; b) data of the partially
    coherent vortex beam in Eq.~(5). Green squares enclose the data
    used for the reconstruction.  The intensity from the central
    microlens vanishes due to the presence of a phase singularity.}
  \label{figData}
\end{figure}

\begin{figure}
  \centerline{\includegraphics[width=0.90\columnwidth]{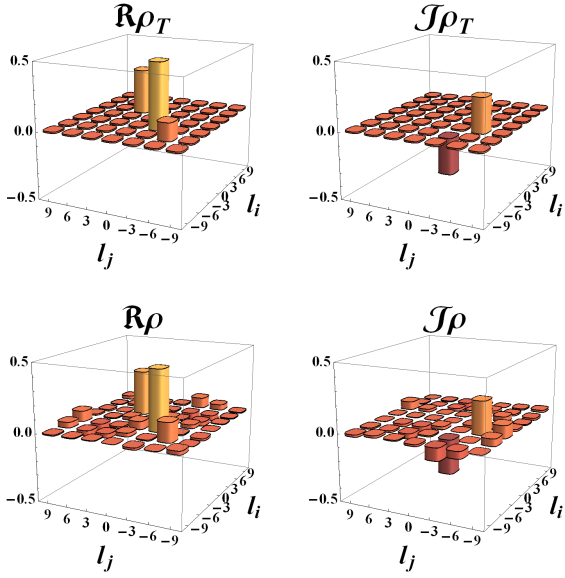}}
  \caption{\textbf{Vortex-beam coherence-matrix reconstruction.} Real
    $\Re$ and imaginary $\Im$ parts of the coherence matrix for the
    true state $\varrho_{\mathrm{true}}$ (upper panel) and for the
    reconstructed $\varrho$ (lower panel).  The reconstruction space
    is spanned by vortex modes with $ \ell \in \{ -9,-6, -3, 0, +3,
    +6, +9\}$. The nonzero values of $\Im \varrho_{-6,-3}$ and $\Im
    \varrho_{-3,-6}$ describe coherences between the modes
    $|V_{-6}\rangle$ and $|V_{-3}\rangle$ and the phase shift $\pi$
    between them.  The very small values of $ \varrho_{3,-6}$,
    $\varrho_{3,-3}$, $\varrho_{-6,3}$ and $\varrho_{-3,3}$ comes from
    the incoherent mixing of $|V_{3}\rangle$ and $|V_{-3}-\frac{i}{2}
    V_{-6}\rangle$. The fidelity of the reconstructed coherence matrix
    is $F =0.98$.}
  \label{figQ}
\end{figure}

To reconstruct the state we use a maximum likelihood
algorithm~\cite{Hradil:2006il,Rehacek:2009jl}, whose results are
summarized in Fig.~\ref{figQ}. The main features of
$\varrho_{\mathrm{true}}$ are nicely displayed, which is also
confirmed by the high fidelity of the reconstructed state $F(
\varrho_{\mathrm{true}}, \varrho)= \Tr[\sqrt{\sqrt{\varrho}
  \varrho_{\mathrm{true}} \sqrt{\varrho}}]=0.98$. The off-diagonal
elements detect the coherence between modes, whereas the diagonal
ones give the amplitude ratios between them. The reconstruction
errors are mainly due to the difference between the true and the
actually generated state.

To our best knowledge, this is the first experimental measurement of
the coherence properties with a wavefront sensor.  The procedure
outperforms the standard SH operation, both in terms of dynamical
range and resolution, even for fully coherent beams. For example, the
high-order vortex beams with strongly helical wavefronts are very
difficult to analyze with the standard wavefront sensors, while they
pose no difficulty for our proposed approach.

The dynamical range and the resolution of the SH tomography are
delimited by the choice of the search space $\{ |k\rangle\}$ and can
be quantified by the singular spectrum ~\cite{Bogdanov:2011fk} of the
measurement matrix $\Pi_{ij}$. For the data in Fig.~\ref{figQ}, the
singular spectrum (which is the analog of the modulation transfer
function in wave optics) is shown in Fig.~\ref{FigSingular}.
Depending on the threshold, around 20 out of the total of 49 modes
spanning the space of $7 \times 7$ coherence matrices can be
discriminated. The modes outside this field of view are mainly those
with significant intensity contributions out of the rectangular
regions of the CCD sensor. Further improvements can be expected by
exploiting the full CCD area and/or using a CCD camera with more
resolution, at the expense of more computational resources for data
post-processing.

\bigskip

\noindent{\textbf{3D Imaging.}}
Once the feasibility of the SH tomography has been proven, we
illustrate its utility with an experimental demonstration of 3D
imaging (or digital propagation) of partially coherent fields.

As it is well known~\cite{Goodman:2005qa}, the knowledge of the
transverse intensity distribution at an input plane is, in general,
not sufficient for calculating the transverse profile at other output
plane.  Propagation requires the explicit form of the mutual coherence
function $G_{\mathrm{in}}$ at the input to determine
$I_{\mathrm{out}}$:
\begin{equation}
  \label{propag}
  I_{\mathrm{out}} ( \mathbf{x} ) = \iint_{-\infty}^{\infty}
  h(\mathbf{x}, \mathbf{x}^{\prime}) 
  h^\ast(\mathbf{x}, \mathbf{x}^{\prime \prime} ) \, 
  G_{\mathrm{in}} (\mathbf{x}^{\prime} , \mathbf{x}^{\prime \prime} )  
  d\mathbf{x}^{\prime} d\mathbf{x}^{\prime \prime}  \, .
\end{equation}
Here $\mathbf{x}^{\prime}$ ($\mathbf{x}^{\prime \prime}$) and $
\mathbf{x}$ are the coordinates parametrizing the input and output
planes, respectively, and $h(\mathbf{x}, \mathbf{x}^{\prime})$ the
response function accounting for propagation.

The dependence of the far-field intensity on the beam coherence
properties is evidenced in Fig.~\ref{figModes} for coherent, partially
coherent and incoherent superpositions of vortex beams.

\begin{figure}
  \centering \includegraphics[width=0.8\columnwidth]{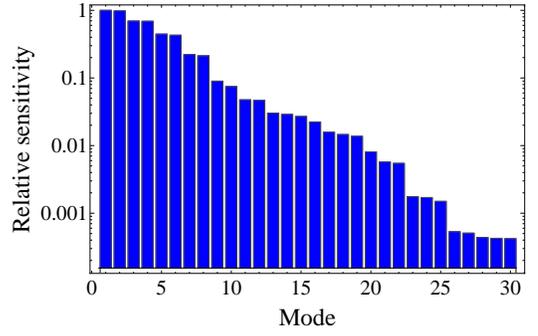}
  \caption{\textbf{Dynamical range of the SH reconstruction.} The
    singular spectrum $\{S_{kk}\}$ of the data in Fig.~4 (here, sorted
    and normalized to the largest singular value) quantifies the
    sensitivity of the tomography setup to the normal modes of the
    problem (see Methods). The relative strengths of the singular
    values correspond to the relative measuring accuracy of those
    modes.  The dynamical range (or field of view) can be defined as
    the set of normal modes with singular values exceeding a given
    threshold.}
  \label{FigSingular}
\end{figure}

\begin{figure}[b]
  \centering
  \includegraphics[width=1.\columnwidth]{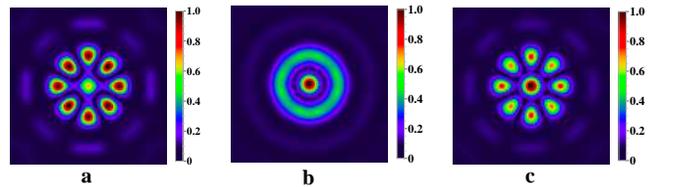}
  \caption{\textbf{Influence of the spatial coherence on the far-field
      intensity distribution.}  We have considered different mixtures
    of the modes $|V_{4}\rangle$, $|V_{-4}\rangle$, and
    $|V_{0}\rangle$ and calculated the associated intensity
    distribution as a Fraunhofer diffraction pattern. (a) fully
    coherent superposition $|V_{4}+V_{-4}+0.4 V_{0}\rangle\langle
    V_{4}+V_{-4}+0.4 V_{0}|$; (b) incoherent mixture
    $|V_{4}\rangle\langle V_{4}|+|V_{-4}\rangle\langle V_{-4}|+0.4
    |V_{0}\rangle\langle V_{0}|$; and (c) partially coherent mixture
    $|V_{4}+V_{-4}\rangle\langle V_{4} + V_{-4}|+0.4
    |V_0\rangle\langle V_0|$.}
  \label{figModes}
\end{figure}

\begin{figure*}
  \centering
  \includegraphics[scale=0.52]{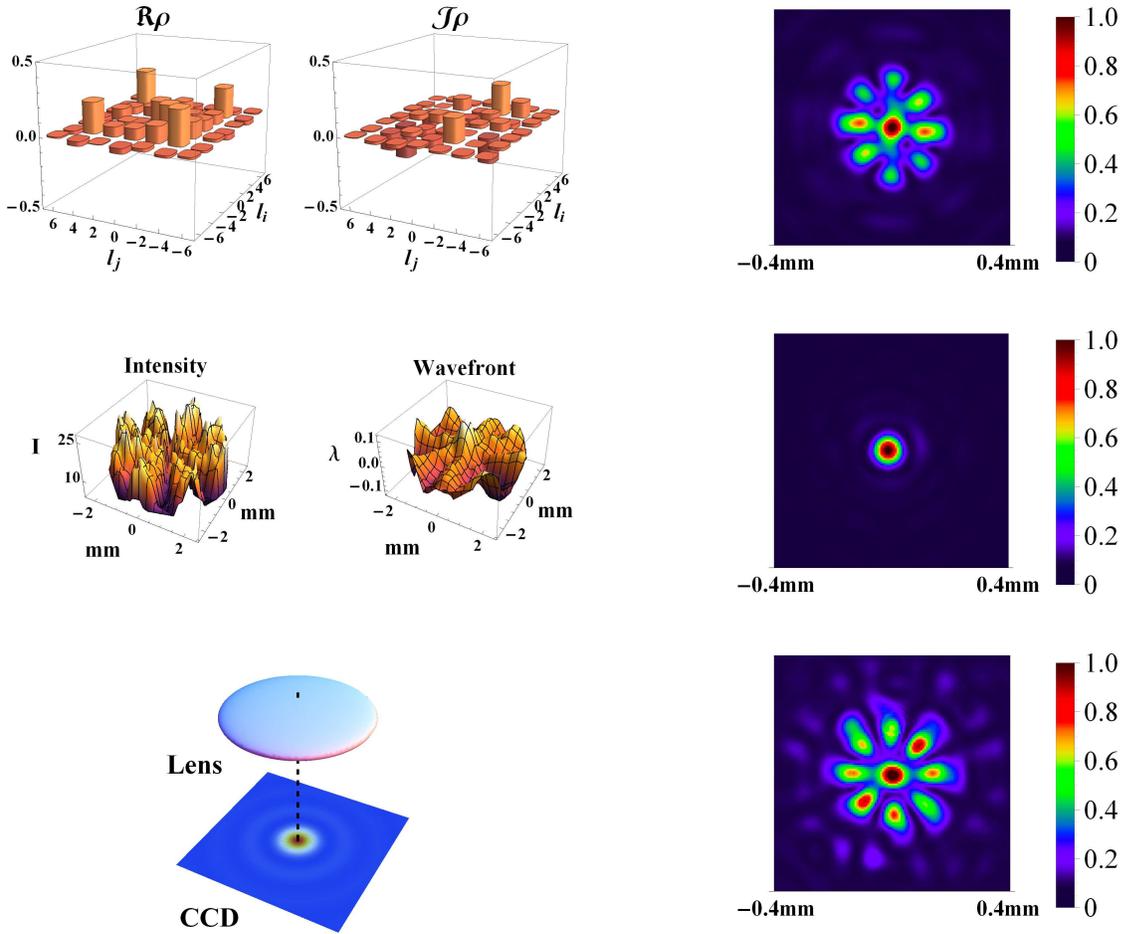}
\caption{\textbf{Digital 3D imaging.}  The prediction of the far-field
  intensity distribution is compared  
with  a  direct intensity measurement. The partially coherent vortex
beam  $|V_{4}+V_{-4}\rangle\langle V_{4} + V_{-4}|+k
|V_0\rangle\langle V_0|$ was generated (with a beam diameter of
4.9~mm) with  a fixed parameter $k$ (unknown prior to the reconstruction).  .
Upper, middle and lower pannels correspond to the SH tomography,
standard SH measurement and direct intensity measurement, respectively.
Upper left: Real and imaginary parts of the reconstructed $\varrho$ in
the 7-dimensional space spanned by the vortices $V_{\ell}$ with $\ell
\in \{-6,-4, -2, 0 , 2, 4, 6\}$.   Upper right: Calculated far-field
intensity distribution $I_{\varrho}$ based on the reconstructed
$\varrho$ propagated to the focal plane of the lens
($f=500$~mm). Middle left: Intensity  distribution (in arbitrary
units) and wavefront as measured by the standard SH sensor.
Middle right: Calculated far-field intensity distribution
$I_{\mathrm{std}}$  using  the standard SH wavefront reconstruction
and the transport of intensity equation included in the sensor
(HASO{\texttrademark}). Bottom left: Schematic picture of the direct
intensity measurement at the lens focal plane. Bottom right: The
result of the direct intensity measurement  $I_{\mathrm{CCD}}$ at the
focal plane with a CCD camera.  
\label{figFarField}}
\end{figure*}

Once the coherence matrix is reconstructed, the forward/backward
spatial propagation can be obtained using tools of diffraction theory
and, consequently, the full 3D spatial intensity distribution can be
computed.  In particular, the intensity profile at the focal plane of
an imaging system can be predicted from the SH measurements. This has
been experimentally confirmed, as sketched in Fig.~\ref{figFarField}.
We prepared the partially coherent superposition
$|V_{4}+V_{-4}\rangle\langle V_{4} + V_{-4}|+k |V_0\rangle\langle
V_0|$, and characterized by the SH tomography method. The
reconstructed coherence function (upper left) was digitally propagated
to the focal plane of a lens and the intensity distribution at this
plane was calculated (upper right) and compared with the actual CCD
scan in the same plane (lower right). Excellent agreement between the
predicted and measured distributions was found.

We emphasize that the standard SH operation fails in this kind of
application~\cite{Schafer:2002rm}. Indeed, we measured the intensity
and wavefront of the target vortex superposition with a standard SH
sensor (middle left) and propagated the measured intensity to the
focal plane using the transport of intensity
equation~\cite{Teague:1983lq,Roddier:1990ai} (middle right). To
quantify the result, we compute the normalized correlation coefficient
[$C(I_{a},I_{b})= \sum_{i,j} I_{a} I_{b} /\sqrt{\sum_{i,j} I_{a}^2}
\sqrt{\sum_{i,j} I_{b}^2}$] of the measured intensity with the
prediction: the result, $C(I_{\mathrm{std}},I_{\mathrm{CCD}})=0.47$,
confirms the inability of the standard SH to cope with the coherence
properties of the signal. This has to be compared with the result for
the SH tomography: $C(I_{\varrho}, I_{\mathrm{CCD}})=0.89$, which
supports its advantages.

\medskip

\noindent{\textbf{Discussion}}

We have demonstrated a nontrivial coherence measurement with a SH
sensor. This goes further the standard analysis and constitutes a
substantial leap ahead that might trigger potential applications in
many areas. Such a breakthrough would not have been possible without
reinterpreting the SH operation as a simultaneous unsharp measurement
of position and momentum. This immediately allows one to set a
fundamental limit in the experimental accuracy~\cite{Appleby:1998fp}.

Moreover, although the SH has been the thread for our discussion, it
is not difficult to extend the treatment to other wavefront sensors.
For example, let us consider the recent results for temperature
deviations of the cosmic microwave
background~\cite{Hinshaw:2009qq}. The anisotropy is mapped as spots on
the sphere, representing the distribution of directions of the
incoming radiation. To get access to the position distribution, the
detector has to be moved and, in principle, such a scanning brings
information about the position and direction simultaneously: the
position of the measured signal prior to detection is
delimited by the scanning aperture, whereas the direction the
signal comes from is revealed by the detector placed at the focal
plane. When the aperture moves, it scans the field repeatedly at
different positions. This could be an excellent chance to investigate
the coherence properties of the relict radiation. To our best
knowledge, this question has not been posed yet.  Quantum tomography
is especially germane for this task.

Finally, let us stress that classical estimation theory has been
already applied to the raw SH image data, offering an improved
accuracy, but at greater computational
cost~\cite{Cannon:1995qf,Barrett:2007ao}. However, the protocol used
here can be implemented in a very easy, compact way, without any
numerical burden.

\bigskip 

\noindent{\textbf{Methods}}

\small
\noindent
\textbf{Partially-coherent beam preparation.}  
Two independent vortex beams were created in the setup of
Fig.~\ref{figSetup} with two laser sources of nearly the same
wavelength: a He-Ne (633~nm) and a diode laser (635~nm). The output
beams were spatialy filtered by coupling them into single-mode fibers.
The power ratio between the modes was controlled by changing the
coupling efficiency. The resulting modes were transformed into vortex
beams by different methods.

The state $ |V_{-3} - \textstyle{\frac{i}{2}} V_{-6} \rangle$ was
realized using a digital hologram prepared with an amplitude spatial
light modulator (OPTO SLM), with a resolution of 1024$\times$768
pixels. The hologram was then illuminated by a reference plane wave
produced by placing the output of a single-mode fiber at the focal
plane of a collimating lens. The diffraction spectrum involves several
orders, of which only one contains useful information.  To filter out
the unwanted orders, a 4$f$ optical processor, with a 0.3~mm circular
aperture stop placed at the rear focal plane of the second lens, was
used. The resulting coherent vortex beam is then realized at the 
focal plane of the third lens.

The second beam $|V_{3}\rangle$ was obtained trough a plane-wave phase
profile modulation by a special vortex phase mask (RPC Photonics).
Finally, the field in Eq.~(5) was prepared by mixing the two vortex
modes in a beam splitter.

During the state preparation, special care was taken to reduce any
deviation between the true and target states. This involved minimizing
aberrations as well as imperfections of the spatial light modulator,
resulting in distortions of the transmitted wavefront.

\bigskip

\noindent 
\textbf{SH detection.}  
The SH measurement involved a Flexible Optical array of 128
microlenses arranged in a hexagonal pattern. Each microlens has a
focal length of 17.9~mm and a hexagonal aperture of 0.3~mm. The signal
at the focal plane of the array is detected by a uEye CCD camera with
a resolution of 640$\times$480 pixels, each pixel being
9.9~$\mu$m$\times$9.9~$\mu$m in size.  Because of microlens array
imperfections, CCD-microlens misalignment, and aberrations of the 4$f$
processor (aberrations of the collimating optics are negligible),
calibration of the detector must be carried out. The holographic part
of the setup provided this calibration wave. SH data from the
calibration wave and the partially coherent beam are shown in
Fig.~\ref{figData}. The beam axis position in the microlens array
coordinates was adjusted with a Gaussian mode.  The detection noise
is mainly due to the background light, which is filtered out prior to
reconstruction.

\bigskip

\noindent 
\textbf{Reconstruction.} The reconstruction was done in the
$7$-dimensional space spanned by the $V_{\ell}$ modes with $ \ell \in
\{ -9,-6, -3, 0, +3, +6, +9 \}$.  All in all, $49$ real parameters had
to be reconstructed. The data come from CCD areas belonging to 7
microlenses around the beam axis; each one of them comprise
11$\times$11 pixels, which means 847 data samples altogether. An
iterative maximum-likelihood
algorithm~\cite{Hradil:2006il,Rehacek:2009jl} was applied to estimate
the true coherence matrix of the signal.

\bigskip

\noindent 
\textbf{Dynamical range and resolution.}  The errors of the SH
tomography can be quantified by evaluating the covariances of the
parameters of the reconstructed coherence matrix $\varrho$.  In the
absence of systematic errors, the Cram\'er-Rao lower
bound~\cite{Cramer:1946ye,Rao:1973qo} can be employed to that end. In
practice, a simpler approach based on the singular spectrum
analysis~\cite{Bogdanov:2011fk} works pretty well.

Let us decompose the $d \times d$ coherence matrix $\varrho$ ($d$ is
just the dimension of the search space) and the measurement operators
$\Pi_{ij}$ in an orthonormal matrix basis $\Gamma_{k}$ ($k=1, \ldots,
d^2$) [$\Tr ( \Gamma_{k} \Gamma_{l} )=\delta_{kl}$], namely
\begin{equation}
  \varrho = \sum r_{k} \Gamma_{k},  
  \qquad 
  \Pi_{ij}=\sum_k p^{ij}_k\Gamma_k,
\end{equation}
so that the Born rule~\eqref{born} can be recast as a system of
linear equations
\begin{equation}
  \label{linsys}
  I_{ij}= \sum_k p^{ij}_k r_k \, .
\end{equation}
Upon using a single index $\alpha$ to label all possible
microlens/CCD-pixel combinations $\alpha \equiv \{i,j\}$,
Eq.~(\ref{linsys}) can be concisely expressed in the matrix form
\begin{equation}
  \label{system}
  \mathbf{I}= \mathbf{P}  \, \mathbf{r} \, ,
\end{equation}
where $\mathbf{I}$ is the vector of measured data, $\mathbf{r}$ is the
vector of coherence-matrix parameters and $\mathrm{P}_{\alpha
  k}=p^\alpha_k$ is the tomography matrix.

Obviously, for ill-conditioned measurements, the reconstruction errors
will be larger and \textit{vice versa}.  By applying a singular value
decomposition to the measurement matrix $\mathbf{P}=\mathbf{U}\,
\mathbf{S}\, \mathbf{V}^\dagger$, Eq.~\eqref{system} takes the diagonal form
\begin{equation}
  \label{systemdiag}
  \mathbf{I}^{\prime}= \mathbf{S}  \, \mathbf{r}^{\prime} \, ,
\end{equation}
where $\mathbf{r}^{\prime}=\mathbf{V}^\dagger \mathbf{r}$ and
$\mathbf{I}^{\prime}=\mathbf{U}^\dagger \mathbf{I}$ are the normal
modes of the problem and the corresponding transformed data,
respectively.  The singular values $\mathrm{S_{kk}}$ are the
eigenvalues associated with the normal modes, so the relative
sensitivity of the tomography to different normal modes is given by
the relative sizes of the corresponding singular values.  With the
help of Eqs.~\eqref{system} and \eqref{systemdiag}, the errors are
readily propagated form the detection $\mathbf{I}$ to the
reconstruction $\mathbf{r}$.

Drawing an analogy between Eq.~\eqref{systemdiag} and
the filtering by a linear spatially invariant system, the singular
spectrum $\mathrm{S}_{kk}$ and the sum of the singular values $\sum_k
\mathrm{S}_{kk}$ are the discrete analogs of the modulation transfer
function and the maximum of the point spread function,
respectively. Hence we define the dynamical range (or field of view)
of the SH tomography as the set of normal modes with singular values
exceeding a given threshold. The sum of the singular values then
describes the overall performance of the SH tomography setup. When
some of the singular values are zero, the tomography is not
informationally complete and the search space must be readjusted.

\bigskip

\noindent
\textbf{Far-field intensity.}  In the experiment on 3D imaging, the
partially coherent vortex beam $|V_{4}+V_{-4}\rangle\langle V_{4} +
V_{-4}|+k |V_0\rangle\langle V_0|$ was generated, where $k$ was a
parameter governing the degree of spatial coherence. To this end, a
coherent mixture $|V_{4}+V_{-4}\rangle\langle V_{4} + V_{-4}|$ was
realized by the digital-holography part of the setup, whereas the
zero-order vortex beam $|V_{0}\rangle$ was prepared by removing the
spiral phase mask. The output diameter of the beam was set to 4.9~mm.

The measurement was done in three steps. First, the SH sensor (see
Fig.~\ref{figSetup}) was replaced by a lens of 500~mm focal length and
the far-field intensity was detected at its rear focal plane with a
CCD camera (Olympus F-View~II, 1376$\times$1032 pixels,
6.45~$\mu$m$\times$6.45~$\mu$m each). Second, the same vortex
superposition was subject to the SH tomography using the SH sensor
(Flexible Optical) and the reconstruction of the coherence matrix in
the 7-dimensional subspace spaned by the vortices $V_\ell$ with $\ell
\in \{ -6,-4, -2, 0 , +2, +4, +6 \}$. Once $\varrho$ is reconstructed,
the far-field intensity was computed using Eq.~\eqref{propag}, where
the focusing is described by the Fraunhofer diffraction response
function.  The predicted intensity was found to be in an excellent
agreement with the direct sampling by the Olympus CCD camera.
Finally, the Flexible Optical SH sensor was replaced by a HASO3 SH
detector. The intensity and wavefront of the prepared vortex beam was
measured and the far-field intensity was computed by resorting to the
transport of intensity performed by the HASO software. Resampling was
done to match the resolution of the HASO output to the resolution of
the Olympus CCD camera.

\bigskip

\noindent
\normalsize{
\textbf{Acknowledgments}}
\small

This work was supported by the Technology Agency of the Czech Republic
(Grant TE01020229), the Czech Ministry of Industry and Trade (Grant
FR-TI1/364), the IGA Projects of the Palack\'y University (Grants
PRF\underline\;\underline\;2012\underline\;\underline\;005 and
PRF\underline\;\underline\;2013\underline\;\underline\;019) and the
Spanish MINECO (Grant FIS2011-26786).

\bigskip

\noindent
\normalsize{ \textbf{Contributions}} 
\small 

The experiment was conceived by J.R., Z.H. and L.L.S.S and carried
out by B.S.  The numerical reconstruction was performed by B.S.,
whereas J. R. and L. M. provided theoretical analysis. Z.H. and
J.R. supervised the project.  The theoretical part of manuscript was
written by L.L.S.S. and J.R. and the experimental part by B. S. and
L.L.S.S., with input and discussions from all other authors.

\newpage


\end{document}